\documentclass{llncs}
\usepackage{amssymb,graphics, amsmath, latexsym}
\usepackage[usenames]{color, epsfig}
\usepackage{subfigure}
\usepackage{hyperref}

\newcommand{\eps}{\epsilon}

\newtheorem{defin}[theorem]{Definition}
\newtheorem{prop}[theorem]{Proposition}
\newtheorem{fact}[theorem]{Fact}

\makeatletter
\def\imod#1{\allowbreak\mkern10mu({\operator@font mod}\,\,#1)}

\renewcommand{\P}{\mathop{\mathrm{P}}}

\begin{document}
\pagestyle{headings}
\title{Approximate Counting of Matchings in Sparse Uniform Hypergraphs}
\author{
  Marek Karpi\'{n}ski\thanks{Research supported partly by DFG grants
      and the Hausdorff Center grant EXC59-1.}\inst{1} and   Andrzej Ruci\'{n}ski\thanks{Research supported by  the Polish NSC grant N201 604 940.}\inst{2} and Edyta Szyma\'{n}ska\thanks{Research supported by the Polish NSC grant N206 565
740.}\inst{2}}

\institute{Department of Computer Science, University of Bonn.
    Email: marek@cs.uni-bonn.de\and
Faculty of Mathematics and Computer Science,
    Adam Mickiewicz University, Pozna\'{n}. Email: rucinski,edka@amu.edu.pl}

\date{}
\maketitle

\begin{abstract}
In this paper we give a fully polynomial randomized approximation scheme (FPRAS) for the number of
 matchings in $k$-uniform hypergraphs whose intersection graphs contain few claws.
 Our method
 gives a generalization of the canonical path method of Jerrum
 and Sinclair to hypergraphs satisfying a local restriction.
 Our proof method depends on an application of
         the Euler tour technique for the canonical
         paths of the underlying Markov chains.
On the other hand, we prove that it is NP-hard to approximate the number of matchings even for the
class of $k$-uniform, 2-regular and linear hypergraphs, for all $k\ge6$, without the above restriction.
\end{abstract}

\newpage

\section{Introduction}\label{intro}

 \emph{A hypergraph} $H=(V,E)$ is a finite set of vertices $V$ together with a family $E$ of
distinct, nonempty subsets of vertices called edges. In this paper we consider $k$-\emph{uniform
hypergraphs} (called also \emph{$k$-graphs}) in which, for a fixed $k\ge2$,  each edge is of size $k$.   \emph{A matching} in a hypergraph
is a set (possibly empty) of disjoint edges.
We will often identify a matching $M$ with the hypergraph $H[M]=(V(M),M)$ induced by $M$ in $H$, where $V(M)=\bigcup_{e\in M}e$.
We denote by $\Delta(H)$  the maximum vertex degree $deg_H(v)$, that is, the maximum number of edges of $H$ containing a vertex $v$.
 A hypergraph is called \emph{linear} (a.k.a. simple)  when no two edges share more than one vertex, that is,
the maximum pair degree is one.

\emph{The intersection graph}  of a hypergraph $H$ is the graph $L:=L(H)$ with vertex set
$V(L)=E(H)$ and  edge set $E(L)$ consisting of all intersecting pairs of edges of $H$. When $H$ is
a graph, the intersection graph $L(H)$ is called \emph{the line graph} of $H$. Every graph $G$ is
the intersection graph of some hypergraph, in fact, of the dual hypergraph $G^*$ of $G$ (obtained
by interchanging the roles of the vertices and edges of $G$, equivalently, by taking the transpose
of the incidence matrix of $G$).

In a seminal paper \cite{js}, Jerrum and Sinclair constructed an FPRAS (see Section \ref{1.1} for
the definition) for counting the number of  matchings in a graph (the monomer-dimer
problem) based on an ingenious technique of canonical paths.
The method was extended later in \cite{JSV} to solve the permanent problem.

Here we  modify their method to address the corresponding problem for $k$-graphs, $k\ge3$. It turns
out that for $k$-graphs $H$, one can
 adopt the proof of the graph case,  whenever for  every two matchings $M,M'$ in $H$ the intersection graph
$L=L(M\cup M')$ between $M$ and $M'$ satisfies  $\Delta(L)\leq 2.$ This happens if and only
if $H$ contains no \emph{3-comb},  a $k$-graph  consisting of  a matching
$\{e_1,e_2,e_3\}$  and one extra edge $e_4$  such that $|e_4\cap e_i|\ge1$ for $i=1,2,3$. Let us denote by ${\cal H}^k_0$ the family of all $k$-graphs which do not contain a 3-comb, cf.~\cite{KRS2}. In Section \ref{examples} we give a couple of examples of  classes of $k$-graphs which belong
to ${\cal H}^k_0$.

By substantially modifying the canonical path method we are able to construct an FRPAS for a
broader class ${\cal H}^k_s$, $s\ge0$, defined as follows. Call an edge $e\in H$ \emph{wide} if it
intersects a matching in $H$ of size at least three (so, every 3-comb contains a wide edge). The
class  ${\cal H}^k_s$ consists of all $k$-graphs containing at most $s$ wide edges.
 Our main result is the
following hypergraph generalization of the Jerrum-Sinclair
theorem. In fact, they, as well as many other contributors to the
field, considered the edge weighted case (with intensity
$\lambda$), while we, for clarity, assume that the hypergraphs are
unweighted ($\lambda=1$). However, the weighted case can be
handled in a similar manner. Our proof method depends on an
application of
         the Euler tour technique for the canonical
         paths of the underlying Markov chains.

\begin{theorem}\label{main}
For every $k\ge3$ and $s\ge0$ there exists an FPRAS for  the problem of counting all matchings in a $k$-graph $H\in{\cal H}^k_s$.
\end{theorem}

 The proof of Theorem \ref{main} is outlined in Section \ref{markov}. We can characterize family ${\cal H}^k_s$ in terms of the intersection graph
$L(H)$. \emph{A claw} in a graph $G$ is an induced subgraph of $G$
isomorphic to the star $K_{1,3}$. The vertex of degree three in a
claw will be called \emph{the center} of that claw. A $k$-graph
$H\in{\cal H}^k_s$ if and only if the intersection graph $L(H)$ of
$H$ contains at most $s$ centers of claws. In particular,
$H\in{\cal H}^k_0$ if and only if $L(H)$ is \emph{claw-free}.
Every $2$-graph, i.e., every graph, is in~${\cal H}^2_0$. For
$k\ge3$, the requirement that $H\in{\cal H}^k_s$ is more
restrictive and causes the hypergraph to be rather sparse (of size
$O(n^{k-1})$). Nevertheless, as can be seen in the next
subsection, the problem of (exactly) counting matchings in
$k$-graphs belonging to ${\cal H}^k_0$ remains computationally
hard.

\subsection{Approximation Hardness}

In this section we demonstrate that the problem of counting matchings in $k$-graphs belonging to
the family ${\cal H}^k_0$ is still  $\#$P-complete, as well as that it is NP-hard to approximate the
number of matchings already for 2-regular, linear $6$-graphs if no restriction on the number of
3-combs is imposed.

\begin{prop}\label{hardness}
The problem of counting matchings in a $k$-graph  $H\in{\cal H}^k_0$ is $\#$P-complete for every
$k\ge3$.
\end{prop}
\proof We use a reduction from the problem of counting all matchings in bipartite graphs $G=(V,E)$
of  maximum degree at most four, which, by a result of Vadhan \cite{vadhan}   is $\#$P-complete.
For a given bipartite graph $G=(V,E)$ of maximum degree at most four with a bipartition $V=V_1\cup
V_2$ we construct a $k$-graph $H=(V',E')$ from the family ${\cal H}^k_0$ as follows. For every edge
$e\in E$ we add to $V$ additional $k-2$ vertices, so $V'=V\cup \bigcup_{e\in
E}\{v^e_1,v^e_2,\ldots,v^e_{k-2}\}.$ Now, every edge $e=(u,v)\in E$ is replaced by the
corresponding $k$-tuple $(v,v^e_1,v^e_2,\ldots,v^e_{k-2},u).$ Thus $|V'|=|V|+(k-2)|E|$, $|E|=|E'|$
and the resulting $k$-graph $H'=(V',E')$ is linear, $k$-partite, has maximum vertex degree at most
four and, more importantly, does not contain a 3-comb. Moreover, there is a natural one-to-one
correspondence between the matchings in $G$ and the matchings in $H.$ \qed

\begin{prop}\label{Aha}
For every $k\ge6$, unless NP=RP there is no FPRAS for the number of matchings in a 2-regular, linear $k$-graph.
\end{prop}
\proof We use a reduction from the problem of approximating the number of independent sets in a
$k$-regular graph, $k\ge6$, for which it has been recently proved (see \cite{sly},\cite{galanis},
and \cite{ss})  that, unless NP=RP, there is no FPRAS. Any $k$-regular graph $G$ is the
intersection graph of the dual hypergraph $H=G^*$, with vertex set $V(H)=E(G)$ and the edges
$e_v\in H$ being  the sets of edges incident to the same vertex $v\in V(G)$. Thus, the number of
independent sets in $G$ equals the number of matchings in $H$. Moreover, observe that by
construction, $H$ is $k$-uniform, 2-regular, and linear. \qed


The  meaning of Proposition \ref{Aha} is that for $k\ge6$ there is
no hope for an FPRAS for the number of matchings even if the
degrees and co-degrees of $H$ are as small as they can get
(1-regular $k$-graphs are matchings themselves and the problems
become trivial). Instead one has to impose some additional
structural restrictions. Inspired by the canonical method of
Jerrum and Sinclair, we came up with the restriction on the number
of 3-combs. In turn, Proposition \ref{hardness} tells us that even
the assumption of no 3-combs at all preserves the computational
hardness, as the problem of exact counting of matchings remains
$\#$P-complete in a quite narrow subclass of ${\cal H}^k_0$.

\subsection{Motivation from Statistical Physics}

In 1972 Heilmann and Lieb \cite{HL} studied monomer-dimer systems, which in the graph theoretic
language correspond to (weighted) matchings in  graphs. In physical applications these graphs are
typically some (infinite) regular lattices. Dimers represent diatomic molecules which occupy disjoint pairs of adjacent vertices of the lattice and monomers are the remaining vertices.
Heilmann and Lieb proved that the associated Gibbs measure is unique (in other words, there is no phase transition). They did it by proving that the roots of the generating matching polynomial of any  graph are all real, equivalently that the roots of the hard core partition function (independence polynomial) of any line graph  are all real. The latter result was later extended to all claw-free graphs by Chudnovsky and Seymour \cite{CS}. The uniqueness of Gibbs measure on $d$-dimensional lattices was reproved in a slightly stronger form and by a completely different method by van den Berg \cite{VdB}.

 Hypergraphs may be at hand when instead of diatomic molecules
 bigger molecules (polymers) are considered which, again, can
 occupy ``adjacent'', disjoint sets of vertices of a lattice.
 As long as the hypergraph lattice $H$ belongs to the family ${\cal H}^k_0$,
 the intersection graph $L(H)$ is claw-free (because $H$ contains no 3-comb)
 and, by the result of \cite{CS} combined with the proof from \cite{HL} there is no phase transition either.
 However, it is possible to have a phase transition
 for a monomer-trimer system (cf. \cite{Heil}). Interestingly, the example given
 by Heilmann (the decorated, or subdivided, square lattice with hyperedges
 corresponding to the collinear triples with midpoints at the branching points of the original square lattice)
 is a 3-uniform hypergraph containing  3-combs, and thus its intersection graph is \emph{not} claw-free.

\subsection{Related  Results}

Recently, an alternative  approach to constructing counting
schemes for graphs has been developed based on the concept of
spatial correlation decay. This resulted in deterministic fully
polynomial time approximation schemes (FPTAS) for counting
independent sets in graphs with maximum degree at most five (\cite{wei}), counting matchings in graphs of bounded
degree (\cite{bgknt}), and, very recently, counting independent sets in claw-free graphs of bounded degree (\cite{suka}). It is not clear to what extent these methods can
be applied to hypergraphs.

The above mentioned result of Weitz \cite{wei} has been recently complemented by the hardness result for graphs with maximum degree at most six, used in the proof of Proposition \ref{Aha} above. It yields  an FPTAS for counting
matchings in hypergraphs whose intersection graphs have  degree at most five. This is the
case of
the Heilmann lattice described in the previous
subsection (the maximum degree of its intersection graph is three), which, by the way, undermines our temptation to link the absence of phase transition for a hypergraph lattice with the absence of a 3-comb, that is with the claw-freeness of the intersection graph of the lattice.
In turn, an FPTAS for counting independent sets in claw-free graphs of bounded degree implies and FPTAS for counting matchings in hypergraphs $H\in{\cal H}^{(k)}_0$ with bounded degree.

As far as hypergraphs are concerned, the authors of \cite{BDK}
showed that, under certain conditions, the Glauber dynamics for
independent sets in a hypergraph, as well as the Glauber dynamics
for proper colorings of a hypergraph mix rapidly. It is
doubtful, however, if the path coupling technique applied there
can be of any use for the problem of counting matchings in
hypergraphs. Nevertheless, paper \cite{BDK} marks a new line of
research, as there have been only few results (\cite{bdgj},
\cite{bd}) on approximate counting in hypergraphs before. The only
other paper devoted to counting matchings in hypergraphs we are
aware of is \cite{barvinok}, where Barvinok and Samorodnitsky
compute the partition function for matchings in hypergraphs under
some restrictions on the weights of edges. In particular they are
able to distinguish in polynomial time between hypergraphs that
have sufficiently many perfect matchings from hypergraphs that do
not have nearly perfect matchings.

\subsection{Approximate Counting and Uniform Sampling}\label{1.1}

Given $\eps>0$ and $\delta>0$, we say that a random variable $Y$ is an $(\eps,
\delta)$-\emph{approximation} of a constant $C$ if $\P\left(|Y-C|\ge\eps C\right)\le\delta.$
Let $f$ be a function  over a set of input strings $\Sigma^*$.

\begin{defin}\label{fpras}\rm
A randomized algorithm  is called a \emph{fully polynomial randomized approximation scheme (FPRAS)
for $f$} if for every triple $(\eps,\delta, x)$ with $\eps>0,\;\delta>0$, and $x\in \Sigma^*,$ the
algorithm returns an  $(\eps, \delta)$-\emph{approximation} $Y$ of $f(x)$ and runs in time
polynomial in  $1/\eps$, $\log(1/\delta)$, and $|x|$.
\end{defin}

Consider a counting problem, that is,  a problem of computing $f(x)=|\Omega(x)|$, where $\Omega(x)$
is a well defined finite set associated with $x$ (think of the set of all matchings in a
hypergraph). As it turns out (see below), to construct an FPRAS for such a problem it is sufficient
to be able to efficiently sample an element of $\Omega(x)$ almost uniformly at random. To make it
precise, given $\eps>0$, we say that a probability distribution $\P:2^\Omega\to[0,1]$ over a finite
sample space $\Omega$ is \emph{$\eps$-uniform} if for every $S\subseteq \Omega$,
$\left|\P(S)-\frac{|S|}{|\Omega|}\right|\le \eps,$
that is, if the total variation distance, $d_{TV}(\P,\tfrac1{|\Omega|})$, between the two
distributions is bounded by $\epsilon$.

\begin{defin}\label{fpaus}\rm
A randomized algorithm  is called a \emph{fully polynomial almost uniform sampler (FPAUS)}
 for a counting problem $|\Omega(x)|$ if for every
pair $(\eps, x)$ with $\eps>0$ and $x\in \Sigma^*,$ the algorithm samples $\omega\in\Omega$
according to an $\eps$-uniform distribution $\P$ and runs in time polynomial in  $1/\eps$ and
$|x|$.
\end{defin}

It has been proved by Jerrum, Valiant, and Vazirani \cite{JVV} that for a broad class of  counting
problems, called self-reducible, including the matching problem, knowing an FPAUS allows one to
construct an FPRAS. For a  proof in the graph case see Proposition 3.4 in \cite{jerrum-book}. The hypergraph
case follows mutatis mutandis. Thus, the proof of Theorem \ref{main} reduces to constructing an
FPAUS for matchings in $H$.

In fact, this approach has been used for perfect matchings
     in dense graphs already by Broder in
\cite{broder}, and later successfully executed by Jerrum and
Sinclair in \cite{js} by different means. In their version the
main steps of finding an efficient  FPAUS for  matchings in a
graph $H$ were

\begin{itemize}
\item[$\bullet$] a construction of an ergodic time-reversible, symmetric Markov chain $\mathcal{MC}(H)$ whose
 state space $\Omega$ consists of
all matchings in $H$;
 \item[$\bullet$] a proof that $\mathcal{MC}(H)$ is rapidly mixing.

\end{itemize}

\subsection{Rapid Mixing}
Given an arbitrary probability distribution $\P_0$ on the state space $\Omega$, let us define the
mixing time $t_{mix}(\epsilon)$ of a Markov chain $\mathcal{MC}$ as
$$t_{mix}(\epsilon)= \min\{t: d_{TV}({\P}_t,\tfrac1{|\Omega|})\le \epsilon\},$$
where $\P_t$ is the chain's state distribution after $t$ steps, beginning from the initial
distribution $\P_0$.  Recall that if an ergodic time-reversible Markov chain is symmetric, i.e.,
the transition probabilities satisfy $p_{ij}=p_{ji}$ for all $i,j\in \Omega$, then its unique
stationary distribution is uniform (cf. \cite{jerrum-book}). In that case we define the transition graph
$G_{\mathcal{MC}}=G$ of $\mathcal{MC}$ as a  graph on the vertex set $V(G)=\Omega$ and the edge set
$E(G)=\{\{i,j\}: p_{ij}>0\}$.  Note that $G$ is undirected but, possibly, with loops. The pivotal
role in estimating the rate of convergence of $\mathcal{MC}$ to its uniform stationary distribution
is played by an expansion parameter, called \emph{ the conductance } and denoted
 $\Phi(\mathcal{MC})$ which in the symmetric case is  defined by a simplified formula
\begin{equation}\label{cond}
\Phi:=\Phi(\mathcal{MC})= min_{S}
  \frac{\sum\{p_{ij}: ij\in G, i\in S, j\in\Omega\setminus S\}}{|S|},
  \end{equation}
  where here (and below) the minimum is taken over all $S\subseteq \Omega$ with $0<|S|\leq \tfrac 12|\Omega|$.
Indeed,
it follows from Theorem 2.2 in \cite{js} that if $p_{ii}\ge\tfrac12$ for all $i\in \Omega$ then
\begin{equation}\label{lam}
d_{TV}({\P}_t,\tfrac1{|\Omega|})\le|\Omega|^2\left(1-\Phi^2/2\right)^t,
\end{equation}
regardless of the initial distribution ${\P}_0$, and consequently,
\begin{equation}\label{mix}
t_{mix}(\epsilon)\le\frac2{\Phi^2}\left(2\log|\Omega|+\log\epsilon^{-1}\right).
\end{equation}
 Hence, it becomes crucial to estimate the
conductance from below by the reciprocal of a polynomial in the input size. To this end, observe that

\begin{equation}\label{phi}
\Phi(\mathcal{MC})\ge min_{S}
  \frac{p_{\min}|cut(S)|}{|S|},
  \end{equation}
  where $cut(S)$ is the edge-cut of  $G$ defined by $S$, and
  $$p_{\min}=\min\{p_{ij}:\; \{i,j\}\in G,\; i\neq
  j\}.$$
 For Markov chains on matchings of an $n$-vertex $k$-graph $H$, denoted further by $\mathcal{MC}(H)$, to bound $|cut(S)|$,  Jerrum and Sinclair introduced their  method of
 canonical paths which boils down to:

\begin{itemize}
\item[$\bullet$] defining a \emph{canonical path} in $G$ for every pair of matchings $(I,F)$ in $H$;
\item[$\bullet$] bounding  from above the number of canonical paths containing a prescribed
 transition (an edge of $G$) by $poly(n)|\Omega|$.
\end{itemize}

Since every canonical path between a matching in $S$ and a matching in the complement of $S$ must go through an edge of $cut(S)$, we have, for $|S|\leq \tfrac 12|\Omega|$,
\begin{equation}\label{cut}
|cut(S)|\ge\frac{|S|(|\Omega|-|S|)}{poly(n)|\Omega|}\ge\frac{|S|}{2poly(n)}
 \end{equation}
 and, by (\ref{phi}),
\begin{equation}\label{Phi}\Phi(\mathcal{MC}(H))\ge\frac{p_{\min}}{poly(n)}.\end{equation}

\section{The Proof of Theorem \ref{main}}\label{markov}
In this section we first outline a proof of Theorem \ref{main} in its special case $s=0$. This
proof is  similar to the proof from \cite{js}. After that we discuss how this proof can be modified
in order to yield the full generality of our main result. (The details  are deferred to the full
version of the paper.)

We begin by defining  a  Metropolis Markov chain whose states are the matchings of a $k$-graph $H$ and then  show that the chain is rapidly mixing to a uniform stationary distribution, yielding an FPAUS.

\subsection{The Markov Chain}
Given a $k$-graph $H=(V,E)$, $|V|=n$, let $\Omega(H)$ denote the set of all matchings in $H.$ We define a
Markov  chain $\mathcal{MC}(H)=(X_t)_{t=0}^{\infty}$ with state space $\Omega(H)$ as follows. Set
$X_0=\emptyset$ and for $t\ge 0$, let $X_t$ be a matching $M=\{h_1,h_2,\ldots,h_s\}$ in $H$, $0\le s\le n/k$. Choose an
edge $h\in H$ uniformly at random and consider the set  $I_h:=\{i:\,h\cap h_i\neq\emptyset,\,i=1,\ldots,s \}$ of the edges of $M$ intersected by $h$. The following transitions from $X_t$ are allowed in $\mathcal{MC}(H)$:
\begin{itemize}
\item[(-)] if $h\in M$ then $M':=M-h,$

\item[(+)] if $h\notin M$ and $|I_h|= 0$ then
$M':=M+h,$

\item[(+/-)] if $h\notin M$ and $I_h=\{j\}$ then
$M':=M+h-h_j,$

\item[(0)] if $h\notin M$ and $|I_h|\ge2$ then
$M':=M$.

\end{itemize}
Finally, with probability 1/2 set $X_{t+1}:=M'$, else $X_{t+1}:=X_t.$

\begin{fact} The Markov chain $\mathcal{MC}(H)$ is ergodic and symmetric.
\end{fact}
\noindent The above fact implies that $\mathcal{MC}(H)$ converges to a stationary distribution that is
uniform over $\Omega(H)$. Moreover, 
\begin{equation}\label{pmin}
p_{\min}=\min\{P_{M,M'}:\; \{M,M'\}\in G,\; M\neq
  M'\}=\frac1{2|H|}\ge n^{-k}.
  \end{equation}

\subsection{Canonical Paths}

In this section we define canonical paths, a tool  used for
estimating the mixing time of the Markov chain $\mathcal{MC}(H)$
introduced in the previous subsection.

For us,  \emph{a path} is a $k$-graph with edge set $\{e_1,\dots, e_m\}$,  $m\ge1$, where for every
$1\le i<j\le m$, $e_i\cap e_j\neq\emptyset$ if and only if $j=i+1$. If $m\ge3$ and, in addition,
$e_1\cap e_m\neq\emptyset$, then such a $k$-graph will be called \emph{a cycle}.
 (Note that a pair of edges sharing at least two  vertices is a path, not a
cycle.)

Set  $V(H)=\{1,2,\ldots,n\}$ and $\min S=\min\{i: i\in S\}$ for any $S\subseteq V(H)$.
Let $(I,F)$ be an ordered pair of matchings in $\Omega(H)$ (we might think
of them as the initial and the final matching of the canonical path-to-be).  The symmetric
difference $I\oplus F$  is a hypergraph with $\Delta(I\oplus F)\le2$ and, due to the
assumption that $H\in{\cal H}^k_0$, also $\Delta(L(I\oplus F))\le2$, that is, in $I\oplus F$ every edge intersects at most two other edges.
 Hence, each component of $I\oplus F$ is a path or a cycle, in which
  the edges of $I$ alternate with the edges of $F$. In particular, each cycle-component has an even number of edges.

Let us order the components $Q_1,\dots,Q_q$ of $I\oplus F$ so that $\min V(Q_1)<\cdots<\min V(Q_q)$. We construct the canonical path $\gamma(I,F)=
(M_0,\dots,M_t)$ in the transition graph $G$ by setting $M_0=I$ and then modifying the current
matching by transitions (+), (-), or (+/-), while
 traversing  the components $Q_1,\dots, Q_q$ as follows. For the sake of uniqueness of the canonical path, each component will be traversed from a well defined starting point (an edge $e_1$) and in a well defined direction $e_1,e_2,\dots e_s$. Of, course, for a path there are just two starting points (which determine directions), while for a cycle there are $s$ starting points and two directions from each. The particular rules for choosing the starting point and direction are quite arbitrary and do not really matter for us.  Suppose that we have
already constructed matchings $M_0,M_1,\dots, M_{j}$ and traversed so far the components
$Q_1,\dots,Q_{r-1}$.

If $Q_{r}$ is an even path then we assume that $e_1\in F$ (and so $e_s\in I$) and take
$M_{j+1}=M_j+e_1-e_2$, $M_{j+2}=M_{j+1}+e_3-e_4$,..., $M_{j+s/2}=M_{j+s/2-1}+e_{s-1}-e_s$.
If $Q_{r}$ is an odd path then we assume that $\min(e_1\cap e_2)<\min(e_{s-1}\cap e_s)$. If
 $e_1, e_s\in I$ then take $M_{j+1}=M_j-e_1$,
$M_{j+2}=M_{j+1}+e_2-e_3$, $M_{j+3}=M_{j+2}+e_4-e_5$, ...,  $M_{j+(s+1)/2}=M_{j+(s-1)/2}+e_{s-1}-e_s$. If $e_1,e_s\in F$, we apply the
sequence of transitions $M_{j+1}=M_j+e_1-e_2$, $M_{j+2}=M_{j+1}+e_3-e_4$,...,$M_{j+(s-1)/2}=M_{j+(s-3)/2}+e_{s-2}-e_{s-1}$, and $M_{j+(s+1)/2}=M_{j+(s-1)/2}+e_s$.
Finally, if $Q_{r}=(e_1,\dots,e_s)$ is a cycle then we assume that $\min e_1=\min(V(Q_r)\cap V(I))$ and $\min(e_2\cap e_3)>\min(e_{s-1}\cap e_s)$, and
follow the sequence of transitions $M_{j+1}=M_j-e_1$,
$M_{j+2}=M_{j+1}+e_2-e_3$, $M_{j+3}=M_{j+2}+e_4-e_5$, ...,$M_{j+s/2}=M_{j+s/2-1}+e_{s-2}-e_{s-1}$, and
$M_{j+s/2+1}=M_{j+s/2}+e_s$.

We call the component $Q_r$ of $I\oplus F$ \emph{the venue} of the transition $(M_j,M_{j+1})$ (on the canonical path $\gamma(I,F)$) if $M_j\oplus M_{j+1}\subseteq E(Q_r)$.
Note that the obtained sequence $\gamma(I,F)=(M_0,\dots,M_t)$ is unique and satisfies the following properties:
\begin{enumerate}
\item[(a)]  $M_0=I$ and  $M_t=F$,
\item[(b)]   for every $j=0,\dots,t-1$, the pair $\{M_{j},M_{j+1}\}$ is an edge
of the transition graph $G$,
\item[(c)]  for every $j=0,\dots,t$, we have $I\cap F\subseteq M_j\subseteq I\cup F$,
\item[(d)] for every $j=0,\dots,t$, we have $F\cap \bigcup_{i=1}^{r-1}Q_i\subseteq M_j$ and $I\cap \bigcup_{i=r+1}^qQ_i\subseteq M_j$, where $Q_r$ is the venue of $(M_j,M_{j+1})$.

\end{enumerate}

\subsection{Bounding the Cuts}

Fix a transition edge $(M,M')$ in  $G$. Let
$\Pi_{M,M'}=\{(I,F):(M,M')\in\gamma(I,F)\}$ be the set of canonical
paths passing through the transition edge $(M,M').$ Our goal is to show that
\begin{equation}\label{Pi}
|\Pi_{M,M'}|\leq
|\Omega_0(H)|,
\end{equation}  where
$\Omega_0(H)=\{H'\subseteq H:\,\exists e\in H'\mbox{ such that }H'-e\in \Omega(H)\}.$ Note that
$|\Omega_0(H)|\le|\{(M,e)\;:\; M\in \Omega(H),\; e\in H\}|\le  n^k|\Omega(H)|$ and
$\log|\Omega(H)|=O(n\log n)$. Thus, in view of the remarks at the end of Section \ref{intro}, the
estimates (\ref{mix}), (\ref{cut}), (\ref{Phi}), (\ref{pmin}), and (\ref{Pi}) yield  a polynomial
bound on $t_{mix}(\epsilon)$ and thus complete the proof of Theorem \ref{main} for $s=0$.

We will prove (\ref{Pi}) by defining a function $\eta_{M,M'}:\Pi_{M,M'}\to \Omega_0(H)$ and
showing that $\eta_{M,M'}$ is an injection. Fix $(I,F)\in \Pi_{M,M'}$ and define
\begin{equation}\label{eta}
\eta_{M,M'}(I,F)=(I\oplus F)\oplus(M\cup M').
\end{equation}

\begin{fact} For all $I,F\in\Pi_{M,M'}$ we have $\eta_{M,M'}(I,F)\in \Omega_0(H).$
\end{fact}

\begin{fact}\label{injection} The mapping $\eta_{M,M'}:\Pi_{M,M'}\to \Omega_0(H)$ is injective.
\end{fact}

\subsection{The General Case}
When 3-combs, or wide edges to that matter, are possible, the structure of a union of two matchings $I$ and $F$ can be much more
complex, as $L(I\oplus F)$ may have vertices of degrees up to~$k$. Nevertheless we are still able
to apply a modification of the canonical path method. For the same Markov chain $\mathcal{MC}(H)$
as before, let us redefine the canonical path $\gamma(I,F)$ as follows. We again order the
components of $I\oplus F$ and focus on a single component $Q_r$. Now, we define a skeleton graph
$S_r$ by replacing each edge of $Q_r$ with a (graph) cycle $C_k$. Note that every vertex of $S_r$
has degree two or four and therefore, by Euler's theorem, there is  an Eulerian tour $E_r$ in
$S_r$. We construct the canonical path $\gamma(I,F)$ in the transition graph $G$ tracing the tours
$E_r$, $r=1,\dots,q$.

First, for every $r$ we select  a start vertex $v_0$ in $E_r$, which is determined by the smallest
indicator. Next, we choose a direction of  each tour in the following way.
\begin{itemize}
\item[(i)]If $\deg_{E_r}(v_0)=4$ then there exist $g\in I$ and
$f\in F$ such that $v_0\in f\cap g.$ Then  the first edge of $E_r$ is $(v_0,w),$ where $w$ is the smaller
of the two neighbors of $v_0$ on $S_r$ which are in $g$ .
 \item[(ii)]If $\deg_{E_r}(v_0)=2$ and there exists $g\in I$ such that $v_0\in g$, then we choose $(v_0,w),$ as above.
 \item[(iii)]If
$\deg_{E_r}(v_0)=2$ and there exists $f\in F$ such that $v_0\in f$, then the first edge of $E_r$ is $(v_0,w),$ where $w$ is the smaller
of the two neighbors of $v_0$ on $S_r$ (which are in $f$).
\end{itemize}

The canonical path  $\gamma(I,F)$ is now being constructed as we follow
 the edges of the Eulerian tours
$E_1,\ldots,E_q$ from the starting points and in the directions defined above. Let us fix
$E_r=(e_1,e_2,\ldots,e_s).$ Suppose that we have traversed already $l-1$ edges of $E_r$
and let $M_{j-1}$ be the current state on the transition path  $\gamma(I,F).$  We have two cases:
\begin{itemize}
\item[1)]if $e_l\subseteq g\in I$ then
if $g\in M_{j-1}$ then $M_j:=M_{j-1}-g$, while if
$g\notin M_{j-1}$ then do nothing;

\item[2)]if $e_l\subseteq f\in F$ then, setting $I_f=\{h_1,\dots,h_m\}$,
if $f\in M_{j-1}$ then do nothing, while if $f\notin
M_{j-1}$ then
$M_j:=M_{j-1}-h_1, M_{j+1}:=M_{j}-h_2,\dots,M_{j+m-2}=M_{j+m-3}-h_{m-1}, M_{j+m-1}=M_{j+m-2}+f-h_m.$
\end{itemize}

So far we have not used the assumption on the bounded number  of wide edges in $H$. But here it
comes. In order to bound $|\Pi_{M,M'}|\leq poly(n)|\Omega(H)|$  we define, as before, the function
$\eta_{M,M'}(I,F)$. However, now $\eta_{M,M'}(I,F)$ is farther away from being a matching. Indeed,
the presence of wide edges may lead to situations where, e.g., $e_1,e_2,e_3\in I$, $e_4\in F$,  and
$e_4\cap e_i\neq\emptyset$, $i=1,2,3$. Then, in the  process of creating the canonical path
$\gamma(I,F)$,  in order to put $e_4$ on the current matching $M_j$ we would need first to delete
$e_1$ and $e_2$, and at least one of them, say $e_2$, by a transition of type (-). As $e_2$ might
intersect two other (than $e_4$) edges of $F$, this may create a path of length three in the set
$\eta_{M,M'}(I,F)$. Fortunately, this scenario can repeat at most $s$ times and, consequently,
$\eta_{M,M'}(I,F)$ belongs to the set
$\Omega_s(H)=\{H'\subseteq H:\,\exists e_0,e_1\dots,e_s\in H'\mbox{ such that }H'-\{e_0,e_1,\dots,e_s\}\in \Omega(H)\}.$
Finally, note that
$|\Omega_s(H)|\le|\{(M,e_0, e_1,\dots,e_s)\;:\; M\in \Omega(H),\; e_0,e_1,\dots,e_s\in H\}|\le  n^{(s+1)k}|\Omega(H)|.$
Theorem \ref{main} follows for any fixed $s\ge0$.

\section{ Hypergraphs with no 3-Combs}\label{examples}

In this section we give a couple of examples of classes of uniform hypergraphs which belong to
family ${\cal H}^k_0$. We concentrate on hypergraphs whose intersection graphs have unbounded maximum degree, so that the result of \cite{suka} does not apply to them.
\subsection{Subdivided 3-graphs}\label{sub}
The following operation generalizes the  edge subdivision in graphs. For an  \emph{arbitrary} 3-graph $H=(V,E)$  construct the \emph{subdivided} 3-graph $H'=(V',E')$ in the following way.
The vertex set is $V'=V\cup V_E$, where $V_E=\{ v_e: e\in E\}$ is disjoint from $V$.
 The edge set $E'$ is obtained by replacing each hyperedge $e=\{v_1,v_2,v_3\}$ with all four triples of the form
 $\{v_i,v_j,v_e\}$.
It is easy to see that for every $H$ the hypergraph $H'$   contains no 3-comb.
Observe that $|H'|=\Theta(|V'|)$ and, depending on the structure of $H$, we might also have $\Delta(L(H'))=\Theta(|V|)$.
Note that for a linear $H$, every matching $M=\{\{u_1,v_1\},\dots, \{u_t,v_t\}\}$ in the shadow
graph $\Gamma(H)$ of $H$ (obtained by replacing each hyperedge with a graph triangle) determines uniquely a matching $M'=\{e_1,\dots,e_t\}$ in
$H'$, where $e_i$ is the unique edge of $H$  containing the pair $\{u_i,v_i\}$. Moreover,
every matching of $H'$ is determined this way. Thus,  for linear $H$, the problem of counting
matchings in $H'$ reduces to counting matchings in graphs.

\subsection{Rooted Blow-up Hypergraphs}\label{root}
Partition an $N$-vertex set $V$ into $n$ nonempty sets $V_1,\dots,V_n$, and fix one vertex $v_i \in
V_i$ for each $i=1,\dots,n$. Fix $k\ge2$ and for every pair $1\le i<j\le n$ include to the edge set $E$ the
family $E_{ij}$ of all $k$-element subsets of $V_i\cup V_j$ containing both, $v_i$ and $v_j$.
Again, it is not hard to see that the obtained  $k$-graph $D=(V,E)$ has no 3-combs.
Note that when $|V_i|=O(1)$ for all $i$, the hypergraph $D$ has $\Theta(n^2)$ edges and $\Delta(L(D))=\Theta(n)$.

 \section{Further Research}\label{fur}

    It remains an open question how to extend
     our result to larger classes of hypergraphs. In particular, in view of Proposition \ref{Aha},
     an intriguing open question is about the existence of an FPRAS for all $k$-uniform hypergraphs,
     $k=3,4,5$. The success  in the case of graphs ($k=2$) relied mostly on the fact that every graph
    is free of 3-combs and thus $I\oplus F$ has a very simple structure. This is the case of the hypergraphs
    in the family ${\cal H}^k_0$ as well. By a more complex argument we were able to prove the existence of
  an FPRAS for ${\cal H}^k_s$, $s\ge0$. For general hypergraphs,
     however, the unlimited presence of wide edges may cause  the image
     of $\eta_{M,M'}$ to become  much larger than $poly(n)\Omega(H)$,  and thus condition (\ref{cut}) might fail.

     Another direction of further research is to try to obtain an FPRAS
    for perfect matchings in \emph{dense} $k$-uniform hypergraphs, where the density is measured as, e.g.,  in \cite{KRS}. For $k=2$ this was done in \cite{js}. The corresponding  decision problem for this class of hypergraphs as well as the problem of constructing a perfect matching was proven in \cite{KRS} to be polynomial time solvable. The 3-combs are an obstacle here too, but in addition, we are facing the problem of the necessity of including into the state space of the Markov chain matchings much smaller than the perfect ones (in \cite{js} the state space consisted only of perfect and near-perfect matchings, that is, matchings missing just two vertices).

\section*{Acknowledgements}

   We thank Martin Dyer, Mark Jerrum and Alex Samorodinsky
    for a number of stimulating discussions.


\newpage



\begin{thebibliography}{HKP99}

\bibitem{barvinok} A. Barvinok, A. Samorodnitsky, Computing the partition function for perfect matchings
 in a hypergraph
 \emph{Combinatorics, Probability and Computing}, 20 (2011), 815--825.

 \bibitem{bgknt}   M. Bayati, D. Gamarnik, D. Katz, C. Nair, and P. Tetali,
    Simple Deterministic Approximation Algorithms for Counting Matchings
\emph{Proc. 39th ACM STOC, San Diego} (2007) 122--127.

\bibitem{BDK} M. Bordewich, M. Dyer, M. Karpinski,
        Path Coupling Using Stopping Times and
        Counting Independent Sets and Coloring in
        Hypergaphs, \emph{Random Struct. Algorithms}
        32 (2008), 375--393.

\bibitem{broder} A. Broder: How hard is it to marry at random?  \emph{Proc. 18th ACM STOC},  (1986) 50--58. (Erratum in \emph{Proc. 20th ACM STOC},
(1988), p. 551)


\bibitem{bdgj}R. Bubley,
  M. Dyer,
  C. Greenhill,
  M. Jerrum: On Approximately Counting Colorings of Small Degree
  Graphs,\emph{SIAM J. Comput.},vol. 29(2), (1999),
  387--400.

\bibitem{bd}R. Bubley, M. Dyer: Graph Orientations with No Sink and an Approximation for a Hard Case of \#SAT,
  \emph{SODA} (1997): 248--257

\bibitem{CS} M. Chudnovsky, P. Seymour, The roots of the independence polynomial of a claw-free graph, \emph{J. Combin. Th., B}, 97 (2007) 350--357.

\bibitem{DFJ} M. Dyer, A. Frieze, M. Jerrum: On Counting
  Independent Sets in Sparse Graphs. \emph{SIAM J. Comput.} 31(5): 1527--1541 (2002).

\bibitem{suka} S. Fadnavis, Approximating Independence
  Polynomials of Claw-Free Graphs, Preprint, Stanford Univ., 2012.

\bibitem{galanis} A. Galanis, Q. Ge, D. Stefankovic, E. Vigoda, L. Yang,
  Improved Inapproximability Results for Counting Independent Sets in the Hard-Core Model,
 \emph{In L. Goldberg, K. Jansen, R. Ravi, and J. Rolim,
   editors, Approximation, Randomization, and Combinatorial Optimization. Algorithms and Techniques,
   volume 6845 of Lecture Notes in Computer Science}, pages 567–578. Springer Berlin / Heidelberg,
 2011. 10.1007/978-3-642-22935-0 48.

\bibitem{Heil} O. J. Heilmann, Existence of phase transitions in certain lattice gases with repulsive potential, \emph{Lettere al Nuovo Cimento} 3 (1972), 95--98.

\bibitem{HL} O.J. Heilmann and E.H. Lieb, Theory of monomer-dimer systems, \emph{Commun. Math. Physics}
  25 (1972), 190--232.

\bibitem{jerrum-book} M. Jerrum: Counting, Sampling and Integrating: algorithms and complexity, Lectures in Mathematics -- ETH Zürich, Birkhäuser, Basel, 2003.

\bibitem{js} M. Jerrum, A. Sinclair: Approximating the Permanent. \emph{SIAM J. Comput.} 18(6): 1149--1178
  (1989).

\bibitem{JSV} M. Jerrum, A. Sinclair, E. Vigoda, A Polynomial-Time
  Approximation Algorithm for the Permanent of a Matrix with
  Nonnegative Entries, \emph{J. ACM} 51 (2004), 671--697.

\bibitem{JVV}     M. Jerrum, L.G. Valiant, V.V. Vazirani, Random Generation of
  Combinatorial Structures from a Uniform Distributions, \emph{Theoret. Computer Sci.} 43 (1986), 169--188.

\bibitem{KRS} M. Karpinski, A. Rucinski, E. Szymanska: Computational Complexity of the Perfect Matching Problem in Hypergraphs with Subcritical Density. \emph{Int. J. Found. Comput. Sci.} 21(6), (2010), 905--924.

\bibitem{KRS2} M. Karpinski, A. Rucinski, E. Szymanska, Approximate Counting of Matchings in Sparse Hypergraphs, arXiv:1202.5885, Feb. 2012.

\bibitem{lv} M. Luby, E. Vigoda: Fast convergence of the Glauber dynamics for sampling independent sets, \emph{Random Struct. Algorithms} 15(3-4): 229--241 (1999).

\bibitem{sly} A. Sly, Computational Transition at the Uniqueness Threshold, \emph{IEEE 51st Annual Symposium on Foundations of Computer Science, FOCS 2010},  pp.287-296, 2010.

\bibitem{ss} A. Sly, N. Sun, The Computational Hardness of Counting in Two-Spin Models
  on Two-Spin Models on d-Regular Graphs, arXiv:1203.2602 v1, 12 March 2012.

\bibitem{vadhan}S. Vadhan: The Complexity of Counting in Sparse, Regular,
  and Planar Graphs, \emph{SIAM J. Comput.} 31(2): 398--427 (2001).

\bibitem{VdB} J. van den Berg, On the Absence of Phase Transition in the Monomer-Dimer Model, \emph{Perplexing problems in Probability (Festschrift in honor of Harry Kesten) Progress in probability}, 44 (1999) 185--195. [ Book chapter ]

\bibitem{wei} D. Weitz, Counting Independent Sets up to the Tree Threshold,
  \emph{Proc. 38th ACM STOC} (2006), 140--149.

\end{thebibliography}
\end{document}